%% file: ms.tex
\documentclass[a4paper, final]{article}
\usepackage{amsmath}
\usepackage{caption}
\usepackage{epigraph}
\DeclareCaptionType{capeq}[figure][List of equations]
\usepackage[dvips,pdftex]{graphicx}
\usepackage{multirow}
\usepackage{booktabs}
\usepackage{url}
\usepackage[all]{xy}
\input{mymath}

\usepackage{natbib}
\usepackage{authblk}
\title{Confounding caused by causal-effect covariability}
\author{Anders Ledberg\thanks{anders.ledberg@su.se or anders.ledberg@gmail.com}}
\affil{Department of Public Health Sciences, Stockholm University SE-106 91 Stockholm, Sweden}
\date{}
\begin{document}
\maketitle

\epigraph{\textit{Any adjustment is somewhat dependent upon the appropriateness of the underlying model -- if the model is appropriate the confounding effect of the prior variables is reduced or eliminated, while if the model is inappropriate a confounding effect remains.}}{-- Donald B. Rubin}

\begin{abstract}
  Confounding seriously impairs our ability to learn about causal relations  from observational data. Confounding can be defined as a statistical association between two variables due to inputs from a common source (the \emph{confounder}). For example, if $Z\rightarrow Y$ and $Z\rightarrow X$, then $X$ and $Y$ will be statistically dependent, even if there are no causal connections between the two. There are several approaches available to \emph{adjust} for confounding, i.e. to remove, or reduce, the association between two variables due to the confounder. Common adjustment techniques include stratifying the analysis on the confounder, and including confounders as covariates in regression models. Most adjustments rely on the assumption that the causal effects of confounders, on different variables, do not co-vary. For example, if the causal effect of $Z$ on $X$ and the causal effect of $Z$ on $Y$ co-vary between observational units, a confounding effect remains after adjustment for $Z$. This \emph{causal-effect covariability} and its consequences is the topic of this paper.

Causal-effect covariability is first explicated using the framework of structural causal models. Using this framework it is easy to show that causal-effect covariability generally leads to confounding that cannot be adjusted for by standard methods. Evidence from data indicates that the confounding introduced by causal-effect covariability might be a real concern in applied work.
\end{abstract}

\section{Introduction}
A core focus in epidemiological studies is on estimating effects of causes. The causes are often called \emph{treatments} or \emph{exposures}  and the effects of interest are typically restricted to some pre-specified \emph{outcomes} \citep[e.g.][]{Rothman-etal2008}. Epidemiological investigations of causal mechanisms are typically formulated in terms of potential outcomes or counterfactuals \citep[e.g][]{Rubin1974, Robins1986}, see \citet{Hernan&Robins2018} for review. According to this framework, the causal effect of a treatment is defined as the difference between the outcome when taking the treatment and the outcome when not taking the treatment. Since only one of these events can be observed (the other is counterfactual), causal effects are typically estimated using observations obtained in different observational units (e.g. participants in a study), where different groups of units receive different treatments. If suitable conditions apply, average causal effects can be estimated consistently and without bias both in experimental and non-experimental settings \citep[e.g.][]{Hernan&Robins2018}.

In most cases the effects of a given cause will vary between observational units -- the treatment might have slightly different effects in different participants. Indeed, this is why the focus typically is on \emph{average} causal effects and not on the causal effects within a particular unit. Variability in the effects of a cause is sometimes referred to as treatment heterogeneity, or causal-effect heterogeneity \citep[e.g.][]{Morgan&Winship2007}, but will here be referred to as \emph{causal-effect variability}, or \emph{effect variability} for short. In experimental studies, where treatments can be randomly assigned to units, average causal effects can be identified in the presence of causal-effect variability. The randomization is expected to cancel out all effect differences between groups receiving different treatments, except those effect differences caused by the treatment \citep[e.g.][]{Rubin1978,Greenland&Robins1986}. In observational studies, on the other hand, the groups exposed to different treatments might also differ in other aspects, and the estimated effects of treatment might therefore be \emph{confounded} with other causal effects \citep{Miettinen&Cook1981,Greenland&Robins1986,Greenland-etal1999,Greenland&Morgenstern2001}. A central theme in observational studies is therefore to find conditions under which groups receiving different levels of treatment are \emph{exchangeable} with respect to the counterfactual outcomes \citep[e.g.][]{Greenland&Robins1986}, see \citet{Hernan&Robins2018} for a general discussion of exchangeability and confounding. In applications, workers try to achieve exchangeability by adjusting for covariates believed to be confounders \citep[e.g][]{Rosenbaum&Rubin1983,Hernan&Robins2018}; where a confounder can be defined as a variable that has a (perhaps indirect) causal effect on both exposure and outcome \citep[see][Ch.~7]{Hernan&Robins2018}. Indeed, properly accounting for confounding is a fundamental requirement for the identification of causal effects of exposures. 

Variability is also expected in the causal effects of confounders. Since confounders, by definition, affects both exposures and outcomes, perhaps through different mechanisms, it is important in this case to consider \emph{causal-effect covariability}. Effect covariability is defined as the extent to which the effects of a given cause co-varies between observational units. In the work presented here, causal-effect covariability of confounders is formalized and the consequences of such covariability are described. In particular, it is shown that when there is causal covariability in the effects of a covariate on exposures and outcomes, residual confounding will remain after adjusting for the covariate. The implication is that statistical associations between exposures and outcomes might be erroneously interpreted as evidence for causal connections between the two. Causal-effect covariability of confounders might be a relatively common phenomenon and deserves more attention in applied work.

In the next section a formal model of causality will be used to study effect covariability of confounders. A simple example will follow, where these consequences are easily seen. Subsequently,  evidence for causal-effect covariability in real data is provided

\section{Theory}
The derivations in this section rely on the structural account of causality as given, for example, in \citet{Spirtes-etal2000} and \citet{Pearl2000}. The counterfactual framework of causality, perhaps more familiar to epidemiologists \citep[e.g.][]{Rubin1974, Robins1986}, can be formulated within the structural account (see the textbooks by \citet{Pearl2000} and \citet{Hernan&Robins2018} for more on the connection between these different frameworks). Some relevant concepts are  introduced below, but for a full account please refer to the references given above.

For clarity, this section will focus on a system with only three observed components, $X$, $Y$, and $Z$ say, where $X$ will stand for exposure, $Y$ for outcome, and $Z$ for confounder.

\subsection{Structural causal models}
In a structural causal model, causal relations are represented by equations which can be mapped onto causal graphs. For example, assume we want to model the dependencies between  $X$, $Y$, and $Z$, with the aim of quantifying the causal effect of $X$ on $Y$. Assume further that the states of $X$ and $Y$ are both partially caused by $Z$ and that $Z$ represents all the common causes of $X$ and $Y$. Other causes (if any) influencing the variables in the model are lumped into ``error terms'' or ``disturbances'' $\varepsilon$. This model is then represented symbolically as 
\begin{align}\label{eq:1}
\begin{split}
  Z &:= \varepsilon_{Z}\\
  X &:= g(Z,\varepsilon_{X})\\
  Y &:= f(X,Z,\varepsilon_{Y}),
\end{split}
\end{align}
where the symbol $:=$ is used to indicate that this system of equations represents a causal identity and not an algebraic one. The functions $g(\cdot)$ and $f(\cdot)$ are non-random, but otherwise general, and the error terms are assumed to be mutually independent. In a structural causal model each equation ``represent a stable and autonomous physical mechanism'' \citep[][p.22]{Pearl2000} implying that it can (at least in principle) be manipulated separately from other equations in the model. This important property of causal models is sometimes referred to as \emph{modularity} \citep[e.g.][]{Woodward2003}.

\subsubsection{Causal diagrams}
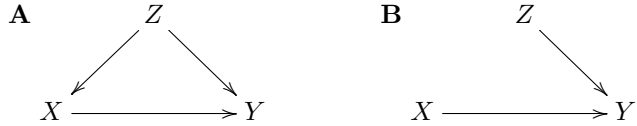
\begin{figure}
  \begin{equation*} 
\mathbf{A}  \xymatrix{& {{Z}}\ar[dr]\ar[dl]& \\
    X \ar[rr]&  & Y}
\quad\quad\quad\quad
\mathbf{B} \xymatrix{& {{Z}}\ar[dr]& \\
  X \ar[rr]&  & Y}
\end{equation*}
\caption[Causal graphs]{Causal graphs. \textbf{A}: causal diagram corresponding to the model of Equation~\ref{eq:1}. \textbf{B}: causal diagram corresponding to the intervened model Equation~\ref{eq:intervention}}\label{graph:1}
\end{figure}

It is often instructive to represent structural causal models as directed acyclic graphs, so called causal diagrams \citep{Pearl1995}. In such graphs the variables in the model correspond to nodes and a direct causal effect between two variables is represented by a directed arrow between corresponding nodes in the graph. In Figure~\ref{graph:1}\textbf{A} the model of Equation~\ref{eq:1} is represented as a causal graph. 

A \emph{path} between two nodes $X$ and $Y$ is a set of non-identical pairs of connected nodes where all nodes, except $X$ and $Y$, occur in exactly two pairs. Thus, starting at $X$, a path is a set of consecutive connected nodes in the graph that leads to $Y$. Often there are several paths joining any two nodes. In Figure~\ref{graph:1} there are two paths between $X$ and $Y$, namely $X\rightarrow Y$ and $X\leftarrow Z \rightarrow Y$. A node $W$ on a path between $X$ and $Y$ that has two incoming arrows (e.g. $X\rightarrow W\leftarrow Y$) is said to be a collider. A path between $X$ and $Y$ is said to be \emph{d-connected} if it does not contain any colliders. A \emph{backdoor path} between $X$ and $Y$ is a d-connected path that starts with an incoming arrow. Thus, the path $X\leftarrow Z \rightarrow Y$ is a backdoor path between $X$ and $Y$. For the work presented here, d-connectedness is important due to the following result: Two variables connected by a d-connected path are statistically dependent, see \citep[][Theorem 1.2.4]{Pearl2000}.

\subsection{Causal effects}
Within the structural framework, causal effects are defined through the notion of an intervention. An intervention can be thought of as a manipulation to force one or more of the variables in the model to take on certain values. Using Pearl's $do()$ notation, the causal effect on $Y$ of $X=x$ (i.e of forcing $X$ to take the value $x$) is written $P(Y|do(X=x))$, and is defined as the probability of $Y$ in the intervened model:
\begin{align}\label{eq:intervention}
\begin{split}
  Z &:= \varepsilon_{Z}\\
  Y &:= f(x,Z,\varepsilon_{Y}).
\end{split}
\end{align}
In words, in the intervened model, the variable $X$ in Equation~\ref{eq:1} is forced to take the value $x$, the rest of the model is left intact (which is possible due to  modularity). In terms of causal diagrams, an intervention in $X$ corresponds to removal of all incoming arrows to $X$ (see Figure~\ref{graph:1}\textbf{B}).

In epidemiology, the causal effects of interest often involve comparisons of two (or more) levels of a treatment. Assuming for the moment that both $X$ and $Y$ are binary indicators of treatment and disease, respectively, then the causal risk difference, as an example, equals
\begin{equation*}
E(Y|do(X=x_1))-E(Y|do(X=x_0)),
\end{equation*}
where expectations are taken with respect to the intervened distributions. 

A fundamental question is if causal effects can be identified from observational data (i.e. without the experimenter being able to manipulate the causal mechanisms). In general this depends on the structure of the causal model \citep[][Ch.3]{Pearl2000}. For the model described by Equation~\ref{eq:1}, the causal effects of $X$ on $Y$ can be identified by adjusting for $Z$. Indeed, if $Z$ is held constant (at $Z=z$ say), then the effect of $X$ on $Y$ equals the causal effect of $X$ on $Y$ (at $Z=z$). Averaging over $Z$ gives the average risk difference in this case:
\begin{align*}
  E&(Y|do(X=x_1))-E(Y|do(X=x_0))=\\
  &\int \left [E(Y|X=x_1,Z=z)-E(Y|X=x_0,Z=z)\right ] dP(z).
\end{align*}
This ``adjustment formula'' is valid because $Z$ ``blocks'' all the backdoor path between $X$ and $Y$ \citep[][Ch.3]{Pearl2000}. 

In the next section it will be shown that the identifiability of the causal effect in this simple model depends critically upon that the causal effects of $Z$ do not co-vary. 
\subsection{Between unit variability}
To proceed we must explicate the concept of an \emph{observational unit}. An observational unit is an entity where quantification (measurement) of variables are made. In medicine and epidemiology this is typically a participant in a study (for example a patient), however, a unit can be some other entity, depending on the scope of the investigation (a hospital, a school, a city, a country, etc). In a typical investigation, observations are made on a number of different units, but there are domains where repeated observations on one unit are more appropriate. 

Note that the causal model described by Equation~\ref{eq:1} contains no reference to observational units. To proceed we therefore make the following 
\begin{assume}
  The structural causal model (Equation~\ref{eq:1}) is an adequate model for all observational units in the population. However, the causal effect of $Z$ on $X$ and $Y$ might differ between units. 
\end{assume}
In other words, we assume that a model of the type represented by Equation~\ref{eq:1}, is a valid approximation of the causal mechanisms \emph{within each person} participating in the study and that the effect of $Z$ (on $Y$ and $X$) might be different in different participants. This a very common (and often implicit) assumption in applied work.

Next we turn to a causal model appropriate for a population of different units. 

\subsubsection{Representing the causal-effect (co-)variability}
Since the focus of this work is on the consequences of causal-effect variability of a confounder (here $Z$), variability in the effect of $X$ on $Y$ will be ignored. Including variability in the causal effect of $X$ will complicate the story, but will not alter the conclusions.

To represent the causal-effect variability of $Z$ in the model we introduce two sets of variables, $U_X$ and $U_Y$ say, and assume that they modulate the effect of $Z$. These sets of variables can be thought of as parameters of the functions $g(\cdot)$ and $f(\cdot)$ respectively. In other words, we write Equation~\ref{eq:1} as 
\begin{align}\label{eq:2}
\begin{split}
  Z &:= \varepsilon_{Z}\\
  X &:= g(Z,U_X,\varepsilon_{X})\\
  Y &:= f(X,Z,U_Y,\varepsilon_{Y}).
\end{split}
\end{align}
In keeping with the assumption above, $U_X$ and $U_Y$ are constant for a particular observational unit, but may vary between units. To extend the model (Equation~\ref{eq:2}) to a causal model for a population of units, we must consider possible causal dependencies between $U_X$ and $U_Y$. One way to make such dependencies explicit is to consider the smallest set of variables, $U$ say, such that $U_X$ and $U_Y$ are conditionally independent given $U$. Or, in terms of the structural model of causality, $U$ is the smallest set that d-separates $U_X$ and $U_Y$ \citep{Pearl2000}. Depending on the set $U$ we get two separate cases. In the first case, $U=\emptyset$, which implies that $U_X$ and $U_Y$ are independent and that their effects therefore can be subsumed into the error terms. In this case there is no causal-effect covariability and Equation~\ref{eq:1} is the appropriate causal model also for the whole population of units. In particular, the average causal effect of $X$ on $Y$ is identifiable also in the population.

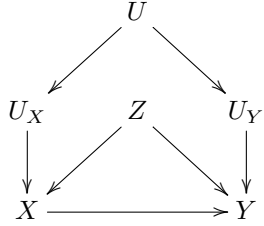
\begin{figure}
\begin{equation*}
  \xymatrix{
    & {{U}}\ar[dr]\ar[dl]& \\
    U_X\ar[d]&Z\ar[dl]\ar[dr]& U_Y\ar[d]\\
   X\ar[rr] &  & Y}
\end{equation*}\caption{Causal graph corresponding to Equation~\ref{eq:full}}\label{fig:full}
\end{figure}

In the second case, when $U\neq \emptyset$, three new equations must be added to the model. A causal model in this case is given by the following equations
\begin{align}\label{eq:full}
\begin{split}
  Z &:= \varepsilon_{Z}\\
  U &:= \varepsilon_{U}\\
  U_X &:= k(U,\varepsilon_{U_X})\\
  U_Y &:= h(U,\varepsilon_{U_Y})\\
  X &:= g(Z,U_Y,\varepsilon_{X})\\
  Y &:= f(X,Z,U_X,\varepsilon_{Y}),
\end{split}
\end{align}
where as before the error terms $\varepsilon_\cdot$ are assumed to be mutually independent. In words, in different observational units the variables in $U$ take different values. The functions $h(\cdot)$ and $k(\cdot)$, transform $U$ to the parameters needed to model the responses to $Z$. The system in Equation~\ref{eq:full} is modular and is a causal model appropriate for a population of units. This model is represented by the causal graph shown in Figure~\ref{fig:full}.
This graph shows that there is a d-connected path between $X$ and $Y$, through $U$, that is not blocked by $Z$. This implies that $X$ and $Y$ are statistically associated within levels of $Z$, independent of the existence of a causal connection from $X$ to $Y$. In particular, we have that
\begin{align*}
  P(Y|do(X=x),Z=z)\neq P(Y|X=x,Z=z).
\end{align*}
Consequently, unless we observe $U$, the causal effect of $X$ on $Y$ is not identifiable.

In the causal model for an individual unit, i.e. Equation~\ref{eq:1}, the only backdoor path between $X$ and $Y$ goes through $Z$. The backdoor path through $U$ in Figure~\ref{fig:full} is a consequence of the (assumed) causal-effect covariability of $Z$. This is stated as a general result.
\begin{result}
Assume that the causal effects of a confounder $Z$ on exposure $X$ and outcome $Y$ varies between observational units. If the effects of $Z$ on $X$ and $Y$ are statistically dependent, then the causal effect of $X$ on $Y$ cannot be identified without adjusting for additional variables.
\end{result}

This result is not just a restatement of a well-known pitfall of observational studies: ``there could always be unmeasured confounders explaining the association between exposure and outcome''. According to the above result, residual confounding \emph{is to be expected} after adjustment for any confounder with covariability in the effects. Thus, unless there are good reasons to believe that the effects of a confounder are either constant or do not co-vary, residual confounding should be considered when interpreting an observed association between exposure and outcome \citep[c.f.][Sec. 7.4]{Hernan&Robins2018}.

\section{An illustrating example}
Here we consider a simple example of a binary exposure $X$, a binary disease indicator $Y$, and a binary confounder $Z$. Assume that the exposure does not cause the disease, and that the confounder takes the values zero and one with equal probabilities (i.e. $P(Z=1) = P(Z=0)$). Assume further that in a given individual, indexed by $i$, $Z$ has the following effect on $X$ and $Y$,
\begin{align*}
  P_i(X=1)&=0.5+\alpha_i(1+Z)\\
  P_i(Y=1)&=0.1+\alpha_i(1+Z),
\end{align*}
where $\alpha_i$ is a parameter. We assume that $\alpha_i$ takes on only two different values, $0.1$ and $0.2$ say, and that these values occur with the same frequency in the study population. For a given observational unit, the effect of $Z$ is consequently an increase in the probability of both exposure and disease with the same amount (since $\alpha$ is constant within units). For example, an individual for which $\alpha=0.1$ would have $P(X=1)=0.6$ and $P(Y=1)=0.2$ if $Z=0$ and $P(X=1)=0.7$ and $P(Y=1)=0.3$ if $Z=1$. Table~\ref{tab:1} shows the probabilities for the different possible cases, illustrating  that $Z$ is increasing the probability of both exposure and outcome. Note that $X$ and $Y$ are independent within levels of $Z$ and $\alpha$. For example, if $\alpha=0.1$ and $Z=0$ then the risk of developing the disease for the exposed is $P(Y=1|X=1)/P(Y=0|X=1)=0.4$; for the unexposed it is $P(Y=1|X=0)/P(Y=0|X=0)=0.4$. Consequently, the relative risk (i.e. $\frac{\text{risk in exposed}}{\text{risk in unexposed}}$) equals 1. 
\begin{table}
  \begin{tabular}{ccccccc}
    & & \multicolumn{5}{c}{$\alpha=0.1$}\\
    \cmidrule{4-6}
    & &\multicolumn{2}{c}{$Z=0$} & &  \multicolumn{2}{c}{$Z=1$}\\
    & & $Y=0$ & $Y=1$ & & $Y=0$ & $Y=1$\\
    \cmidrule{3-4}
    \cmidrule{6-7}
    & $X=0$ & 0.32 & 0.08 & & 0.21 & 0.09 \\
    & $X=1$ & 0.48 & 0.12 & & 0.49 & 0.21 \\
    &  &  &  & &  &  \\
    &  &  &  & &  &  \\
    & & \multicolumn{5}{c}{$\alpha=0.2$}\\
    \cmidrule{4-6}
    & &\multicolumn{2}{c}{$Z=0$} & &  \multicolumn{2}{c}{$Z=1$}\\
    & & $Y=0$ & $Y=1$ & & $Y=0$ & $Y=1$\\
    \cmidrule{3-4}
    \cmidrule{6-7}
    & $X=0$ & 0.21 & 0.09 & & 0.05 & 0.05 \\
    & $X=1$ & 0.49 & 0.21 & & 0.45 & 0.45 \\
  \end{tabular}
  \caption{Probabilities for the example, normalized within levels of $Z$ and $\alpha$, i.e. $P(X,Y|Z,\alpha)$.}\label{tab:1}
\end{table}

However, assume now that the between-unit variability in the effects of $Z$ are not observed (i.e. assume $\alpha$ is not observed). Using the assumption of equal proportion of individuals with $\alpha=0.1$ and $\alpha=0.2$, the observable probabilities in this case are those shown in Table~\ref{tab:2}. The relative risk of getting the disease is now associated with exposure, within both levels of $Z$. Indeed, for $Z=0$ the relative risk is $1.08$ and for $Z=1$ it is $1.3$, corresponding to an average relative risk of $1.19$. The marginal (over levels of $Z$) relative risk is $1.34$, so stratifying the analysis on $Z$ reduces the ``effect size'' but leaves a substantial residual due to variability in the causal effect of $Z$. Thus,  after adjusting for $Z$, the relative risk is still confounded by the effects of $Z$ (the true (causal) relative risk is 1.0, by assumption), and this residual confounding is due to causal-effect covariability. 
\begin{table}
  \begin{tabular}{ccccccc}
    & &\multicolumn{2}{c}{$Z=0$} & &  \multicolumn{2}{c}{$Z=1$}\\
    & & $Y=0$ & $Y=1$ & & $Y=0$ & $Y=1$\\
    \cmidrule{3-4}
    \cmidrule{6-7}
    & $X=0$ & 0.26 & 0.08 & & 0.13 & 0.07 \\
    & $X=1$ & 0.48 & 0.16 & & 0.47 & 0.33 \\
  \end{tabular}
  \caption{Probabilities for the example, normalized within levels of $Z$ ($P(X,Y|Z)$), and rounded.}\label{tab:2}
\end{table}

The assumption that the effect modulator $\alpha$ only could take two different values is inessential for the effect to appear. Similar effects are obtained if $\alpha$ varies randomly between units. In other words, if the exposure and outcome probabilities in each individual would have the following form
\begin{align*}
  P_i(X=1)&=0.5+\alpha_{iX}(1+Z)\\
  P_i(Y=1)&=0.1+\alpha_{iY}(1+Z)
\end{align*}
it would suffice that $(\alpha_{iX},\alpha_{iY})$ are statistically dependent for an effect similar to that shown in Table~\ref{tab:2} would occur. 

In the next section, evidence of causal-effect covariability in real data will be demonstrated.

\section{Causal-effect covariability in real data}
Here I will describe an observational data-set and show that a probability model that includes causal-effect covariability fits these data better than a model that does not. The primary aim is to provide an example of causal-effect covariability in real data, not to describe a scientifically relevant case that warrants further study. The example might appear remote from typical studies in epidemiology, but should not be taken to imply that causal-effect covariability is a minor issue in such studies. The implication is rather that there is a lack of public data-sets where a convincing demonstration of causal-effect covariability can be made. 
\subsection{The variables}
To build on the formalism of previous sections we need a three-variable system where it is reasonably likely that two of the variables ($X$ and $Y$, say) are not causally affecting each other, but are mutually affected by the third variable, $Z$. Furthermore, as shown in the Appendix, if the causal effects of $Z$ on $X$ and $Y$ are approximately linear, then the presence of causal-effect covariability implies that the covariance of $X$ and $Y$, conditional on $Z$, will have a particular dependence on $Z$. Thus, a linear causal effect of $Z$ leads to constraints that can be used to fit a probability model to the data.

The following three variables, arguably, approximately fulfills the above conditions: The variable $X$ will stand for strength of the knee extensor muscles (knee strength henceforth), and $Y$ for hand-grip strength (grip strength). Body weight will serve as the confounder ($Z$). Note that body weight is a variable that can (in principle) be manipulated within individuals and that it therefore makes sense to conceive of causal models in each observational unit.

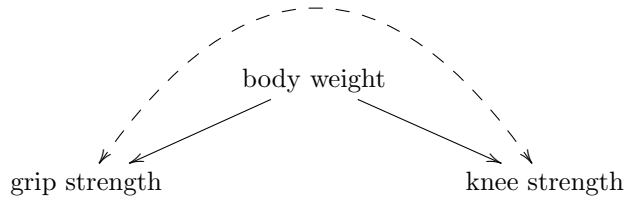
\begin{figure} 
\begin{equation*}
  \xymatrix{& {\mathrm{body\ weight}}\ar[dr]\ar[dl]& \\
    \textrm{grip strength}\ar@/^5.5pc/@{<-->}[rr] &  & \textrm{knee strength}}
\end{equation*}\caption{Tentative causal graph for the real data case. The dashed line represent unmeasured confounding.}\label{fig:exp}
\end{figure}

A change in body weight is likely to cause a change in muscle strength, proportional to the change in weight, for the following two reasons: (i) if weight increases, more force is required to move around, and the extra force is proportional (i.e. linear) to the increase in weight. (ii) Bigger muscles weigh more and the strength of a muscle is roughly proportional to the cross sectional area of the muscle \citep{Maughan-etal1983}, which is proportional to the weight of the muscle. It is also reasonable to assume that strength of leg and arm muscles do not exert a direct causal influence on each other. There are, of course, many other causes of knee strength and grip strength and some of these causes will affect both variables (e.g. physical exercise). Figure~\ref{fig:exp} shows a tentative causal diagram for the causal effects between the variables in a single individual, the dashed line represents unmeasured confounders. 
\subsection{A probability model for the data}

The following probability model will be used to fit the data for each level of the confounder $Z$:
\begin{align}\label{eq:linear}
\begin{split}
  X&=\mu_X+\beta_XZ+\varepsilon_X\\
  Y&=\mu_Y+\beta_YZ+\varepsilon_Y\\
  \beta_X&=b_X+\varepsilon_{\beta_X}\\
  \beta_Y&=b_Y+\varepsilon_{\beta_Y}
\end{split}
\end{align}
with
\begin{equation*}
  (\varepsilon_{\beta_X},\varepsilon_X,\varepsilon_{\beta_Y},\varepsilon_Y)\disteq N(\mat{0},\smat{\Sigma}),
\end{equation*}
where $N(\mat{0},\smat{\Sigma})$ represents the multivariate normal distribution with zero mean and covariance matrix $\smat{\Sigma}$, and $\disteq $ denotes equality in distribution. The key term in this model is the covariance between the coefficients of $Z$, i.e. between $\varepsilon_{\beta_X}$ and $\varepsilon_{\beta_Y}$. This term models the causal-effect covariability of $Z$. To test the contribution of this term, the model represented by Equation~\ref{eq:linear} was fit both with and without the inclusion of this term. The models were fit by numerical maximization of the likelihood. See the Appendix for more details.  

\subsection{The data}
Data on these three variables were obtained from a large public-domain data set originating from measurements made on young Swedish men tested as part of enlistment for compulsory military service. The data-set contains observations of a large number of variables in men who were tested during the years 1969-1996. Most men were around 18-21 years of age when tested. Select parts of this data set has featured in a large number of research publications. Both hand grip and knee extensor strength were measured using highly standardized protocols, the details of which are unfortunately not public \citep{Ortega-etal2012}. Data and R-code is available in the following GitHub repository: \url{https://github.com/aledberg/causal-covar}.

\subsubsection{Case selection}
To make a linear relation between body weight and muscle strength more plausible (this is needed to model causal-effect covariability) the range of body weights were restricted to be between 64 and 75 kg. Persons with weights outside this interval were excluded. To further increase the plausibility of a linear causal model, persons not having a body height in the interval 178-181 cm were excluded (height might also have a causal effect on muscle strength, and this restriction will lessen the impact of height). After these steps there remained observations from approximately 175000 persons in the data set. The data can be thought of as representing hypothetical changes of the weight-strength relationship in persons of very similar heights with an initial weight of $69$ kg (the average weight in this population).\footnote{Of course, weight can be perturbed in many different ways, and the causal effect on strength will likely depend on the perturbation used (see \citet{Hernan&Taubman2008} for a related discussion on using weight as an intervention).} 

\subsection{Results}
\begin{figure}
  \centerline{
    \includegraphics[scale=.5]{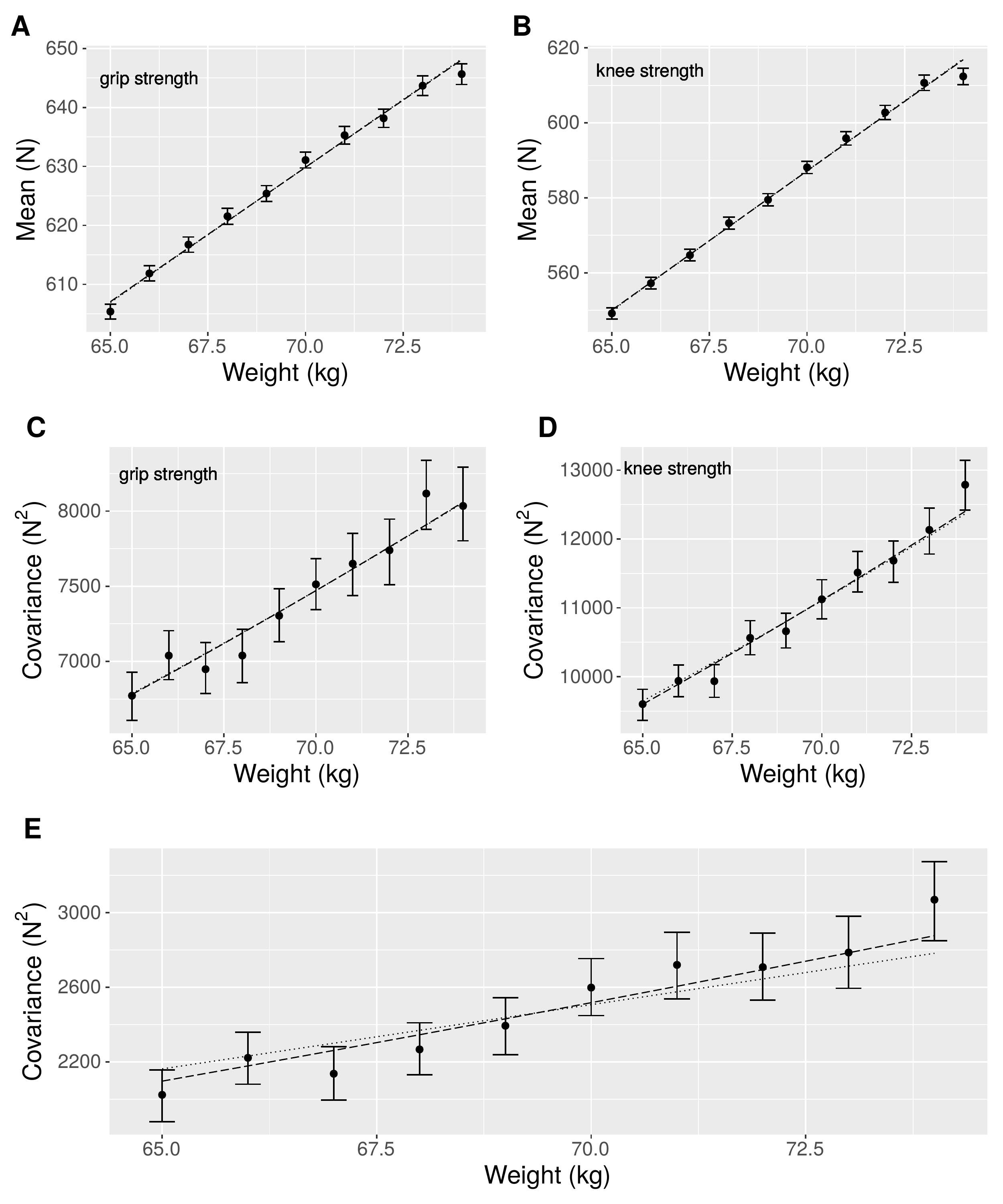}}
  \caption[Conditional means, variances and covariance.]{
    Conditional means, variances and covariance for the real data. Dashed lines show model predictions for the ``full'' model (with 13 parameters). Dotted lines show predictions based on the model assuming no causal-effect covariability. Error bars show approximate 95\% confidence intervals. \textbf{A}: Conditional mean for grip strength. \textbf{B}: Conditional mean for knee strength. \textbf{C}: Conditional variance for grip strength, confidence intervals estimated by bootstrapping. \textbf{D}: Conditional variance for knee strength, confidence intervals estimated by bootstrapping. \textbf{E}: Conditional covariance of grip strength and knee strength, confidence intervals estimated by bootstrapping.
  }
  \label{fig:1}
\end{figure}

Figures~\ref{fig:1}A and \ref{fig:1}B show that both mean hand grip and mean knee strength increase approximately linearly with body weight. The data were selected to show such linear dependency, and the linear relation breaks down if a wider range of weights are considered (not shown). Figures~\ref{fig:1}C and \ref{fig:1}D show that the variance of grip and knee strength depend on body weight, an indication of effect variability. Figure~\ref{fig:1}E shows that the covariance between grip and knee strength also depends on body weight, an indication of effect covariability.

To test the importance of causal-effect covariability in explaining the data, two versions of the probability model (Equation~\ref{eq:linear}) were fit to the data. A version containing the parameter representing effect covariability (full model), and a version without that parameter (reduced model). The results of the fits are shown as dashed lines (full model) and dotted lines (reduced model) in Figure~\ref{fig:1}. Visual inspection indicates that the model which included causal-effect covariability better predicted the conditional covariance. A more quantitative comparison is obtained by comparing the likelihoods of the fits. This show that the full model explained more of the variability in the data (likelihood-ratio test, D=13.65, $p=2.2 \cdot 10^{-4}$). Table~\ref{tab:fit} shows the parameter values for the two models. A simple calculation, using the parameters in the table and the expression for the conditional covariance given in the appendix,  shows that the covariability of $\beta_X$ and $\beta_Y$ accounted for most of the linear correlation between knee and hand grip strength. The contributions of the regression terms, $b_X$ and $b_Y$, were minor in comparison. 
\begin{table}
  \begin{tabular}{r|cc|cc}
    & \multicolumn{2}{c}{full model} & \multicolumn{2}{c}{reduced model} \\
    & & & &  \\
    parameter & estimate & 95\% C.I. & estimate & 95\% C.I \\
    \hline
     $m_X$ & 67.7 & (52.6, 116.3) & 65.0& (54.0, 72.5)\\
     $m_Y$ & 310.9 & (279.4, 332.2)& 308.6 & (305.9, 309,8)\\
     $b_X$ & 7.4 & (6.7,7.6)& 7.5& (7.34, 7.61)\\
     $b_Y$ & 4.6 & (4.2,5.0)& 4.6& (4.56, 4.63)\\
     $\sigma^2_{\varepsilon_X}$& 114.4 & (37.9, 437.0)& 3213 & (3172, 3249)\\
     $\sigma^2_{\varepsilon_Y}$& 2401 & (1169, 5475)& 2575 & (2554, 2624)\\
     $\sigma_{\varepsilon_X\varepsilon_Y}$& -523.0 & (-653.9, 679.9) & -2317& (-2332, -2265)\\
     $\sigma^2_{\beta_X}$& 2.24 & (2.03 2.49)& 2.75 & (2.73, 2.76)\\
     $\sigma^2_{\beta_Y}$& 1.01 & (0.23, 1.25) &1.03 &(1.01, 1.04)\\
     $\sigma_{\beta_X\beta_Y}$& 0.63 & (0.25, 0.67) & n.a. & (not in model)\\
     $\sigma_{\beta_X\varepsilon_x}$& 0.25 & (-10.3, 5.3) &-39.77 & (-39.84, -39.64) \\
     $\sigma_{\beta_y\varepsilon_Y}$& 0.81 & (-17.5, 23.4)&-1.05 & (-1.11, -0.95)\\
    $\sigma_{\beta_x\varepsilon_Y}$& -0.23 & (-3.2, 8.9)&34.5 &(34.0, 34.7) \\
    \hline
    log likelihood & \multicolumn{2}{c}{1757909.73} & \multicolumn{2}{c}{1757916.56} \\
  \end{tabular}
  \caption{Maximum likelihood estimates of model parameters. Confidence intervals based on 160 bootstrap resamples for the full model and 500 for the reduced model.}\label{tab:fit}
\end{table}

\section{Discussion}
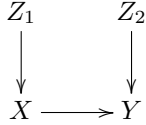
\begin{figure}
  \begin{equation*} 
    \xymatrix{Z_1\ar[d] & Z_2\ar[d] \\
      X \ar[r]&   Y}
  \end{equation*}
  \caption[Causal graph]{Causal graph illustrating the possibility of confounding without a confounder.}\label{graph:discussion}
\end{figure}

In this work I have formalized the notion of causal-effect covariability and shown that if there is covariability in the effects of a confounder, then residual confounding will remain after adjustment for the confounder. Furthermore, evidence in support of causal-effect covariability was demonstrated in real data. In this section some issues regarding the interpretation of the main result are discussed, as are conditions under which causal-effect covariability might be expected in applications. First however, it is important to emphasize that the case of three observed variables used throughout the paper was chosen for didactic reasons. In more realistic models, with many more interacting variables, causal-effect covariability can lead to confounding through many routes. For example, assume that the causal diagram shown in Figure~\ref{graph:discussion} describes the causal mechanisms in a single observational unit. According to this model only the direct causal connection from $X$ to $Y$ will contribute to a statistical dependency between $X$ and $Y$ within a single observational unit. However, if we assume that the effect of $Z_1$ on $X$ and $Z_2$ on $Y$ varies between observational units and moreover that these effects are statistically dependent, then the causal effect of $X$ on $Y$ cannot be identified from observations made on different units without adjusting for other variables in addition to $Z_1$ and $Z_2$. 
In this model causal-effect variability leads to confounding without a confounder! In terms of the structural model of causality, this case corresponds to violation of the modularity assumption (i.e. the error terms are not independent). 

\subsection{Adjusting for the \emph{level} or the \emph{effect} of a confounder?} 
One way of thinking about effect covariability of confounders is to make a distinction between \emph{the level} of a confounder and \emph{the effect} of a confounder. In the example above, the confounder $Z$ had two \emph{levels} (zero or one), but the \emph{effect} of $Z$ was an increase of the probabilities of exposure and outcome by $0.1, 0.2, 0.3$ or $0.4$ units (i.e. four levels). As shown, adjusting for the level of the confounder led to residual confounding. On the other hand, knowing also the effect of the confounder, it was possible to fully adjust for the confounding introduced by $Z$ (for example by stratification, as in Table~\ref{tab:1}).

It is perhaps contrary to common terminology to claim that the statistical association between $X$ and $Y$ is caused by $Z$, when $Z$ is held constant. On the one hand side, when $Z=1$ in the above example, all units receive the same input (by assumption), so the input from $Z$ cannot contribute variability to $X$ or $Y$ (and hence not to their covariability). On the other hand, the \emph{effect} of $Z$ differs between units (since $\alpha$ varies between observational units), so the effect of $Z=1$ \emph{does} contribute variability to $X$ and $Y$. This highlights the difference between adjusting for the level of a variable, and adjusting for the effect that this variable has. Importantly, when the effect of a confounder is constant across observational units, adjusting for the level of the confounder or for the effect of the confounder would both fully account for the confounding introduced by the confounder.

One could perhaps argue that $Z$ and $\alpha$ in the example are two different confounders, and by adjusting for only one of them there will of course be residual confounding. However, this misses the important point that the two confounders are intrinsically connected -- $\alpha$ modulates the effect of $Z$ -- and that these confounders should therefore be considered together.

In applications workers sometimes write that they adjust or control for ``the effect of'' a confounder, when they have in fact adjusted for the level of the confounder\footnote{A search in the Web of Science Core Collection using variants of ``controlling/adjusting for the effect(s) of'' indicate that this is a relatively common misconception as several thousands of research papers make such claims in the abstract. Most of these papers are not in journals focusing on Epidemiology, indicating that this misconception is widespread.}. Perhaps this is an indication of that it is not uncommon to assume constant effects of covariates. Note that the effects of a confounder are typically not observed, and the applied researcher can only hope that adjusting for the levels of a confounder will be sufficient to remove all the confounding introduced. It is a real possibility that the magnitude of confounding introduced by effect covariability could exceed the magnitude of confounding introduced by the different levels of the confounder. This seemed to be the case in the real data analyzed here.

\subsection{Between or within unit variability}
In this work the focus has been on between-unit variability of causal effects, and it was assumed that the causal effects were constant within a given observational unit. However, this is not a necessary condition for residual confounding to occur. Indeed, causal-effect variability \emph{within} units might often be enough to introduce residual confounding due to covariability of causal effects. It is important to carefully consider all sources of variability when designing a study, and to make sure that these sources are appropriately modeled in the analysis. 

\subsection{Random effects models}
Causal-effect covariability can also be understood in terms of fixed- and random effects models. If the effects of a confounder are assumed to be fixed (i.e. same in all observational units) then current approaches to adjustment will remove the confounding. If the effects of a confounder are more appropriately modeled as random effects (i.e. the true effect size varies between observational units) then the resulting confounding might not be removed by adjustment, as shown here. In the real data example, the (assumed) linear relation between the weight and the outcome variables (and the large number of observations) enabled an estimate of the random effects of weight. It was shown that the covariability of the effects of weight accounted for most of the statistical association between the outcome variables; the contribution of the fixed effect was minor in comparison. This might not be representative for observational studies in general, but residual confounding through effect covariability might be relatively common.  

Unfortunately, many designs used in observational studies do not allow for estimation of random effects of covariates. If effect covariability is a concern it might therefore be advantageous to use designs were such random effects can be estimated \citep[c.f][]{Greenland2000}. 

\subsection{How common is causal-effect covariability}
That the effects of a given cause varies between observational units is probably very common. However, only when there is covariability between the effects on the exposure and the effects on the outcome will these two be confounded even after adjustment. Indeed, to generate the effect of the confounder $Z$ in the example above (Table~\ref{tab:2}), each individual had the same constant added to the probabilities of exposure and outcome, perhaps not a very likely scenario in the real world.

So how common is causal-effect covariability? This is an empirical question that deserves more attention, but it seems likely that the presence or not of covariability in part depends on the ``distance'' between the mechanisms producing the effects on the exposure and the mechanisms producing the effects on the outcome. Consider the following example: Assume that ``work environment'' (WE) is a potential confounder and consider first a study looking at the effect of smoking on lung cancer. Its reasonable to assume that WE is a proxy for factors that could cause both smoking and lung cancer. However the factors that cause smoking (mainly social factors, say) and those that cause lung cancer (exposure to work-related toxic substances, say) act through very different mechanisms (peer pressure vs mutations in respiratory epithelium cells, say). In this case there are perhaps no \emph{a-priori} reasons to believe that individuals more susceptible to peer pressure would be more or less susceptible to carcinogenic substances. When there is a large ``distance'' between the causal mechanisms, causal-effect covariability might be small. 

Consider now using the same covariate (WE) in a study of causal effects of smoking on heavy drinking (of ethanol-containing beverages). In this case it is possible that similar factors that cause smoking also cause heavy drinking (different social norms in different work places, say). Moreover, in this case the causal mechanisms are probably more similar -- an individual more likely to develop a regular smoking behavior might also be more likely to develop a regular (heavy) drinking behavior (due to similar neural mechanisms involved in learning these behaviors, say). In this case a more substantial causal-effect covariability seem likely, and adjusting for work environment might not completely remove the association between smoking and drinking caused by WE. 

In the real-data example the observations could be explained by a model that included effect covariability. It might be possible to account for the data equally well (similar likelihood) with models not including covariability, a more solid proof of concept would probably require an experimental study. Still, the provided evidence should be enough to warrant further investigations of causal-effect covariability in data. 

\subsection{Conclusions}
Causal-effect covariability of a confounder introduces statistical associations between exposures and outcomes that cannot be controlled for by conventional methods. In many empirical studies effect covariability of confounders cannot be ruled out based on prior scientific knowledge. Effect covariability of confounders should therefore be included as a possible explanation of the association between exposures and outcomes in such studies. 

\section*{Acknowledgments}
This research has been financed by the Swedish Council for Working Life and Social Research (FAS 2006--1523) and by  The Swedish National Board of Institutional Care (grant number: 2.6.1-1134-2017). I thank Daniel Chicharro for invaluable discussions and Rita Almeida for reading and commenting on previous versions of this paper.

\clearpage
\bibliography{refs}

\appendix
\clearpage
\section{Appendix}
\subsection{A linear causal model of the data}
Here we consider a model for the real data described in the main text. In other words, we want to model the effect of the three variables, hand grip strength $X$, knee extension strength $Y$ and body weight $Z$. We assume that there is linear causal relation between $Z$ and $X$ and $Y$. Since there will most likely be other factors influencing $X$ and $Y$ we need to include those in the model as well (even if they are unobserved); let these common factors be represented by $W$. We further assume that the effects of $W$ on $X$ and $Y$ are independent of the effects of $Z$.\footnote{This assumption is made to limit the number of free parameters that must be estimated from data. The assumption can be relaxed at the cost of introducing more variables.}  This gives the following causal model
\begin{align}\label{eq:data}
\begin{split}
  Z &:= \varepsilon_{Z}\\
  W &:= \varepsilon_{W}\\
  U &:= \varepsilon_{U}\\
  (\alpha_X,\beta_X) &:= k(U,\varepsilon_{k})\\
  (\alpha_Y,\beta_Y) &:= h(U,\varepsilon_{h})\\
  X &:= \alpha_X +\beta_X Z + \theta(W)+\widetilde{\varepsilon_X}\\
  Y &:= \alpha_Y+\beta_Y Z + \phi(W)+\widetilde{\varepsilon_Y},
\end{split}
\end{align}
where all the error terms ($\varepsilon$'s) are assumed to be mutually independent. 

To continue it is helpful to re-write some of the terms in the above system as a constant term (not changing between units) and a fluctuating term. So, let $\alpha_X=m_X+\varepsilon_{\alpha_X}$ where $m_X$ represents the population average (over units) and $\varepsilon_{\alpha_X}$ the deviations from this average, and similarly for $\alpha_Y$. Do a similar decomposition of $\beta_X$, i.e. write $\beta_X=b_X+\varepsilon_{\beta_X}$, and similarly for $\beta_Y$. Finally, let $\theta(W)=t_X+\varepsilon_\theta$, and $\phi(W)=t_Y+\varepsilon_\phi$. Next, we simplify the equations for $X$ and $Y$ by defining  new error terms as
\begin{align}\label{eq:error}
  \begin{split}
    \varepsilon_X=&\varepsilon_{\alpha_X} + \varepsilon_\theta+ \widetilde{\varepsilon_X}\\
    \varepsilon_Y=&\varepsilon_{\alpha_Y} + \varepsilon_\phi+ \widetilde{\varepsilon_Y}
  \end{split}
\end{align}
and new constant terms as 
\begin{align}\label{eq:mean}
  \begin{split}
    \mu_X=&m_X+t_X\\
    \mu_Y=&m_Y+t_Y.
  \end{split}
\end{align}

With this reformulations the equations for $X$ and $Y$ become
\begin{align}\label{eq:xy}
\begin{split}
  X &:= \mu_X+ (b_X+\varepsilon_{\beta_X}) Z + \varepsilon_X\\
  Y &:= \mu_Y+ (b_Y+\varepsilon_{\beta_Y}) Z + \varepsilon_Y.
\end{split}
\end{align}
This change in notation simplifies the expressions for the conditional variances and covariances below, but note that the error terms for $X$ and $Y$ (i.e. $\varepsilon_X$ and $\varepsilon_Y$) are no longer independent in this notation. Using standard rules for variances and covariances it is straight forward to derive two consequences of causal effect variability of $Z$:
\begin{constraint}\label{pred:1}
  The conditional variance of $X$ (and $Y$), given $Z$, is a quadratic function of $Z$ and is given by
  \begin{equation}\label{eq:cond_var}
    V(X|Z)=\sigma_{\beta_X}^2Z^2+2\sigma_{\beta_X\varepsilon_X}Z+\sigma_{\varepsilon_X}^2,
  \end{equation}
and similarly for $V(Y|Z)$. 
\end{constraint}
Here $\sigma_{\beta_X}^2$ is the variance of $\varepsilon_{\beta_X}$, $\sigma_{\varepsilon_X}^2$ the variance of $\varepsilon_X$, and $\sigma_{\beta_X\varepsilon_X}$ is the covariance between these terms. This expression shows that if there is causal-effect variability in a linear causal model, the conditional variance will depend on $Z$.

The next constraint involves covariability of the effects of $Z$.
\begin{constraint}\label{pred:2}
  The conditional covariance of $X$ and $Y$, given $Z$, is a quadratic function of $Z$ and is given by
  \begin{equation}\label{eq:cond_covar}
    C(X,Y|Z)=\sigma_{\beta_X\beta_Y}Z^2+(\sigma_{\beta_X\varepsilon_Y}+\sigma_{\beta_Y\varepsilon_X})Z+\sigma_{\varepsilon_X\varepsilon_Y}
  \end{equation}
\end{constraint}
Here $\sigma_{\beta_X\beta_Y}$ is the covariance between $\varepsilon_{\beta_X}$ and $\varepsilon_{\beta_Y}$, $\sigma_{\varepsilon_X\varepsilon_Y}$ the covariance between $\varepsilon_{X}$ and $\varepsilon_{Y}$, $\sigma_{\beta_X\varepsilon_Y}$ the covariance between $\varepsilon_{\beta_X}$ and $\varepsilon_{Y}$, and similarly for $\sigma_{\beta_Y\varepsilon_X}$. Next we use these constraints to derive a probability model for the data. 

\subsection{A probability model for the data}
To probe the evidence for causal-effect covariability in the data a probability model is described here and fit to the data using maximum likelihood. In particular, for a given value of $Z$, we will assume that the following model describes the data:
\begin{align*}
  X&=\mu_X+\beta_XZ+\varepsilon_X\\
  Y&=\mu_Y+\beta_YZ+\varepsilon_Y\\
  \beta_X&=b_X+\varepsilon_{\beta_X}\\
  \beta_Y&=b_Y+\varepsilon_{\beta_Y}
\end{align*}
with
\begin{equation*}
  (\varepsilon_{\beta_X},\varepsilon_X,\varepsilon_{\beta_Y},\varepsilon_Y)\disteq N(\mat{0},\smat{\Sigma}),
\end{equation*}
where $N(\mat{0},\smat{\Sigma})$ represents the multivariate normal distribution, and $\disteq $ denotes equality in distribution. There are 14 parameters in this model (two constant terms, two slope terms, and ten variance-covariance terms). However, the two terms $\sigma_{\beta_X\varepsilon_Y}$ and $\sigma_{\beta_Y\varepsilon_X}$ cannot both be identified given the constraints derived above. We therefore set $\sigma_{\beta_X\varepsilon_Y}=\sigma_{\beta_Y\varepsilon_X}$ in the following. With this additional constraint, the covariance matrix becomes
\begin{equation*}
  \smat{\Sigma}=\begin{pmatrix}
  \sigma^2_{\beta_X}& \sigma_{\beta_X\varepsilon_X} & \sigma_{\beta_X\beta_Y}& \sigma_{\beta_X\varepsilon_Y} \\
  \cdot& \sigma^2_{\varepsilon_X}& \sigma_{\beta_X\varepsilon_Y} & \sigma_{\varepsilon_X\varepsilon_Y} \\
  \cdot & \cdot & \sigma^2_{\beta_Y}& \sigma_{\beta_Y\varepsilon_Y}\\
  \cdot& \cdot & \cdot  & \sigma^2_{\varepsilon_Y} \end{pmatrix}.
\end{equation*}
To investigate the importance of the causal-effect covariability, a model where $\sigma_{\beta_X\beta_Y}$ was forced to be zero was also fit to the data. In this ``reduced model'' the covariance matrix takes the following form
\begin{equation*}
  \smat{\Sigma}=\begin{pmatrix}
  \sigma^2_{\beta_X}& \sigma_{\beta_X\varepsilon_X} & 0& \sigma_{\beta_X\varepsilon_Y} \\
  \cdot& \sigma^2_{\varepsilon_X}& \sigma_{\beta_X\varepsilon_Y} & \sigma_{\varepsilon_X\varepsilon_Y} \\
  \cdot & \cdot & \sigma^2_{\beta_Y}& \sigma_{\beta_Y\varepsilon_Y}\\
  \cdot& \cdot & \cdot  & \sigma^2_{\varepsilon_Y} \end{pmatrix}.
\end{equation*}

The models were fit by maximizing the normal likelihood numerically over levels of $Z$. In other words, for each level of $Z$, the covariance matrix of $X$ and $Y$ is formed using the above expressions for conditional variance and covariance, and the parameter values that maximize  the likelihood are found numerically. Local maxima are abundant as the model is close to be overparameterized, and care was taken to find the global maximum. Standard errors of the fitted parameters were obtained by bootstrapping.  Code to fit the model to data is provided in following GitHub repository: \url{https://github.com/aledberg/causal-covar}. 
\end{document}

%% file: mymath.tex
\usepackage{amsmath}
\usepackage{amsthm}
\usepackage{amsfonts}
\usepackage[mathscr]{euscript}
\usepackage{mathrsfs}

\newcommand{\mat}[1]{\begin{mathbf} {#1} \end{mathbf}}
\newcommand{\smat}[1]{\boldsymbol{#1}}

\newcommand{\disteq}{\stackrel{\mathrm{d}}{=}} 

\theoremstyle{definition}

\theoremstyle{plain}

\newtheorem{constraint}{Constraint}
\theoremstyle{plain}

\newtheorem*{assume}{Assumption}
\newtheorem*{result}{Main result}
\newtheorem*{observation*}{Observation}
\newtheorem*{prediction*}{Prediction}
